\documentclass[twoside,a4paper,11pt]{sca}
\usepackage{graphicx}
\usepackage{hyperref}
\usepackage{movie15}
\usepackage{natbib}  
\begin{document}
\pagenumbering{arabic}
\pagestyle{myheadings}
\thispagestyle{empty}
{\flushright\includegraphics[width=\textwidth,bb=90 650 520 700]{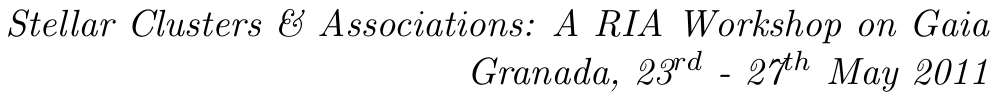}}
\vspace*{0.2cm}
\begin{flushleft}
{\bf {\LARGE
%
Spectroscopic and  photometric surveys  of the Milky Way and
its stellar clusters in the Gaia era
%
}\\
\vspace*{1cm}
%
Sofia Feltzing$^{1}$
%
}\\
\vspace*{0.5cm}
%
$^{1}$
Lund Observatory, Department of Astronomy and Theoretical Physics, Box 43, SE-221 00 Lund, Sweden\\
%
\end{flushleft}
%
\markboth{
Spectroscopic and photometric surveys
}{ 
%
Feltzing, S.
%
}
\thispagestyle{empty}
\vspace*{0.4cm}
\begin{minipage}[l]{0.09\textwidth}
\ 
\end{minipage}
\begin{minipage}[r]{0.9\textwidth}
\vspace{1cm}
\section*{Abstract}{\small
%

This contribution to the {\sl Stellar Clusters \& Associations: 
A RIA Workshop on Gaia} deals with surveys of stars, in 
particular spectroscopic surveys, with 
attention to their impact on cluster studies, and their connection
with {\sl Gaia}. I review some of the scientific reasons why we want 
large spectroscopic surveys, what requirements these put on the 
instrumentation. Then I turn to a review of the
 current and future instrumentation that 
will enable us to complement {\sl Gaia}'s excellent distances and 
proper motions with the desired ground-based spectroscopy to obtain additional
radial velocities and elemental abundances of high quality. 
This is a very fast moving area with several new
surveys using existing and future multi-object spectrographs on 4- and
8-meter telescopes. As things change rapidly it is difficult to give
adequate information on all these projects, but with the inclusion of
links to the relevant web-sites the reader should be able to follow
the latest developments as they unfold.

%
\normalsize}
\end{minipage}
%
%
%
\section{Surveying the Milky Way and its stellar components \label{mw}}

In the past the Milky Way was surveyed to find our place in the
Universe. \citet{1918ApJ....48..154S,1918PASP...30...42S} showed,
using globular clusters, that the Sun's position in the Galaxy is not
in any privileged central part but indeed quite far from the centre,
towards the outer 1/3 of the stellar disk. Later studies focused on
the oldest stars in order to figure out how old the Milky Way and its
stars are. This enabled much work on the stellar populations in the
Milky Way and provided constraints on theories of galaxy formation in
general \citep[e.g.,][]{1962ApJ...136..748E}. A good overview to our
understanding of the Milky Way through large surveys and to the future
prospects at the time is provided in the proceedings of the Joint
Discussion 13 at the 26th IAU General Assembly \citep[][in particular
the contribution by \citet{2006MmSAI..77.1036W}, see the footnote for
where to access the articles]{2006MmSAI..77.1026C}\footnote{The table
  of content and all pdf-files are available at
  \href{http://sait.oat.ts.astro.it/MSAIt770406/index.html}{\tt
    http://sait.oat.ts.astro.it/MSAIt770406/index.html}}.

Today, we survey our Galaxy to find out how the Milky Way fits in the
general framework of galaxy formation and evolution provided by
$\Lambda$CDM
\citep[e.g.,][]{2008MNRAS.391.1685S,2005Natur.435..629S}. Or rather --
we wish to use the Milky Way as a testbed for such models. The
examples of this approach are many. Interesting examples include
\citet{2011MNRAS.tmp..885H}, \citet{2009MNRAS.400L..61S}, and
\citet{2003ApJ...597...21A}. A striking example of the
interplay between observations and theory is provided by the
prediction that the number of, then, known dwarf spheroidal galaxies
(dSph) around the Milky Way was at least an order of magnitude too low
as compared with the then current predictions from the
state-of-the-art $\Lambda$CDM modelling \citep{1999ApJ...524L..19M}.
At the time, about ten dSph galaxies were known
\citep{1998ARA&A..36..435M}.  This lead to much work studying if, for
example, star formation could be inhibited in small dark matter halos,
thus providing a solution such that they would never start shining
with stellar light.  The Sloan Digital Sky Survey
\citep[SDSS,][]{2006ApJS..162...38A} later provided a deep,
multi-band, photometric survey of a significant portion of the
sky. Vasily Belokurov and colleagues analysed this material and
quickly found several more dSph, which were also confirmed
spectroscopically, i.e., that the systems have a common radial
velocity and are not just asterisms on the sky
\citep{2007ApJ...654..897B,2009MNRAS.397.1748B}.  A particularly
difficult example is given by the Hercules dSph
\citep{2007ApJ...654..897B,2009A&A...506.1147A}.  An extrapolation
from these, and subsequent investigations, points to the possibility
that the Milky Way might actually be surrounded by almost as many dSph
as predicted by $\Lambda$CDM. However, along the way many important
lessons about galaxy formation and evolution have been
learnt. \citet{2011arXiv1103.6234G} provides a  concise update on
the theoretical as well as the observational studies tackling
this problem.

\citet{2002ARA&A..40..487F} reviewed the prospects of identifying,
with the help of {\sl chemical tagging}, individual regions of star
formation in the Galaxy. The chemical tagging concept builds on the
understanding that all stars form in clusters and that each cluster
should have a unique signature, not only in age but also in elemental
abundances. The ability to identify unique groups of stars in this way
is challenging, requires very large spectroscopic surveys, and has
lead to the proposal that underpins the HERMES project. We will return
to the topic of chemical tagging and elemental abundances in
Sects.\,\ref{sect:field} and \ref{sect:gaia}.

\section{Stellar clusters}
\label{sect:clusters}

Stellar clusters are an intrinsic part of any galaxy and provide vital
clues to as diverse subjects as star-formation, stellar evolution,
nucleosynthesis, understanding dynamical interactions between stars,
and galactic formation and evolution. In many galaxies they are the
most luminous individual objects that we can study also at very large
distances. They can be used to constrain the formation histories of
galaxies and put constraints on models of galaxy formation and
evolution \citep{2006ARA&A..44..193B,1991ARA&A..29..543H}.  In the
Milky Way, the two major classes of stellar clusters, open and
globular, span a large range of ages and metallicities thus having the
potential to play the role to constrain the formation and evolution of
our Galaxy. The globular cluster population in the Milky Way contains
only old clusters
\citep{2009ApJ...694.1498M,2005AJ....130..116D,1999AJ....118.2306R}.
They divide into mainly two groups as concerns ages -- the largest
subset, containing only old clusters with very little spread in ages
and a younger subset of clusters (a couple of billion years difference
at the most) which are, based on their position within the Galaxy and
chemical and kinematic properties, interpreted as being accreted from
other systems, such as the Sagittarius dSph galaxy
\citep{2010ApJ...718.1128L}.  In other galaxies, e.g., the Large
Magellanic Cloud, there are both old and young globulars
\citep[e.g.,][]{2011ApJ...735...55C}. The open cluster population in
the Milky Way mainly span younger ages, although there are open
clusters as old as 10\,Gyr \citep[see][with its associated
data-base\footnote{The data-base is available, together with links to
  many other open cluster catalogues, at
  \href{http://www.astro.iag.usp.br/~wilton/}{\tt
    http://www.astro.iag.usp.br/$\sim$wilton/}}]{2002A&A...389..871D}. The
globular clusters have a broad range of metallicities with two
distinct peaks \citep[something which is common to many
galaxies,][]{2006ApJ...636...90H}, but they do not cover very
metal-poor stars (below --2.2\,dex) nor stars with solar metallicities
or higher
\citep{1985ApJ...293..424Z,2010arXiv1012.3224H}\footnote{Note that
  this is a new update on the {\sl Catalog of Milky Way Globular
    Clusters} by W.E.~Harris.  The full data-base can be found at
  \href{http://physwww.mcmaster.ca/~harris/Databases.html}{\tt
    http://physwww.mcmaster.ca/$\sim$harris/Databases.html}}.  The
open clusters span metallicities higher than about $-0.5$\,dex
\citep[e.g., the compilation in][and the work by
\citet{2010AJ....139.1942F}]{2009A&A...494...95M}.

The different ranges in metallicities and ages for the two cluster
populations is also reflected in (and connected with?) their spatial
distributions. The globular clusters in the Milky Way mainly reside in
a roughly spherical distribution. \citet{1959MNRAS.119..559K}
indicated that the globular cluster system in the Milky Way is
composed of two kinematically distinct sub-groups with differing
metallicities.  The study by \citet{1985ApJ...293..424Z} then later
clearly showed that the more metal-rich globulars, [Fe/H]$>$--0.8, are
centrally concentrated, whilst the less metal-rich globulars (in fact
the majority of all the globulars in the Milky Way) form a much more
extended, spherical system. These metal-poor globulars are thought to
be associated with the stellar halo, although some of them are likely
accreted from other, smaller, galaxies. Palomar\,12 is one such
example \citep[][]{2000AJ....120.1892D,2010ApJ...718.1128L}.
\citet{2010MNRAS.404.1203F} discuss additional associations of
globulars with dSph galaxies. For the metal-rich globulars
sub-division into bulge and disk clusters have been discussed
\citep[see examples in,
e.g.,][]{1995AJ....109.1663M,1996ASPC...92..211Z,1998ASPC..136...33H}
but such divisions remain an open subject.  For example,
\citet{2006A&A...450..105B} found that the metal-rich globulars have a
spherical distribution, whilst \citet{2003AJ....125.1373D} found that
at least one of them (NGC\,6528) is, based on its kinematics,
associated with the bar. This points to the necessity of getting more
and better kinematic data also for globulars in order to understand
the globular cluster system. In a series of important paper Dana
Casetti-Dinescu and collaborators \citep[][and references
therein]{2010AJ....140.1282C} have painstakingly derived the proper
motions and space velocities for by now 34 globulars using the
Southern Proper Motion Program data, steadily improving our
understanding of the dynamics of the system.  Very few globulars
appear to be genuinely associated with the stellar disks. In fact most
globular clusters are too metal-poor to be associated with the
disks. The so far only, confirmed, disk globular is NGC\,5927 for
which the full 3D kinematic information shows it to have an
essentially disky orbit \citep{2007AJ....134..195C}. The cluster has a
metallicity, [Fe/H]= --0.37, as well as elemental abundances, e.g., Ca
is enhanced relative to Fe, that naturally associate it with the thick
disk (Simmerer et al., a study of 7 HB stars in the cluster, to be
submitted).

Studies of elemental abundances in globular clusters is a large
subject that was recently reviewed in detail by
\citet{2004ARA&A..42..385G}. The globulars do in general follow the
abundance trends of the major stellar populations in the Milky Way,
however, there is increasing evidence that there are aspects of
``clusterness'' that might set their stars apart from the stars
normally observed in the field, e.g., they show Na-O anti-correlations
which are not seen in the field stars \citep[see,
e.g.,][]{2009A&A...505..139C,2009A&A...505..117C,2011ApJ...730L..16M}
and more and more clusters are shown to have multiple stellar
populations also based on their colour-magnitude diagrams
\citep[e.g.,][]{2007ApJ...661L..53P}. The discovery of which has
resulted in much theoretical work and deepening our understanding of star
formation and evolution \citep[see, e.g.,][]{2011MNRAS.tmp..736D}. A
good example of how an existing survey can be used to tackle a new
problem is given by \citet{2011A&A...525A.114L} who used publicly
available photometry from SDSS to trace the two populations (UV-red
and UV-blue) in globular clusters (cores excluded) and find that radial
variations in the populations are present. This is feasible as the
survey covers the cluster outskirts very well, till they dissolve into
the field population, something that is not always feasible with
dedicated programs.

Open clusters is a much larger subject than globular clusters. In
particular since the clusters include young and heavy stars. Thus rare
evolutionary stages can be studied in detail. The reader is directed
to other contributions to this workshop for more details on the
massive and young stars. Spectroscopic surveys will naturally have a
large impact on the study of those rare evolutionary stages, however, I
will limit myself to aspects of the open clusters that mainly concerns
them as birthplaces of field stars, tracers of Galactic chemical
evolution, structure, and dynamics.

In total we know of more than 1800 open clusters detected in the visual
\citep[][and subsequent updates]{2002A&A...389..871D}. For those about
50\% have an age estimate \citep{2010IAUS..266..106M}. In addition
there are about 700 clusters detected in the near infra-red. For the
about 200 open clusters with abundance estimates the most striking
feature is probably the Galactic radial gradient in metallicity that
they trace. Currently, very few other tracers are able to deliver well
determined metallicities combined with good distance estimates (for
the open clusters distances come from isochrone fitting). It appears
that the disk as traced by the open clusters shows a steadily
declining [Fe/H] as a function of galacto-centric distance, until it
reaches a floor at about 10-12\,kpc
\citep{2010AJ....139.1942F,2010sf2a.conf..335B}. Similar observations
are not readily available for large data-sets in the field. A recent
example of a small study of field giants shows, however, the same
floor in metallicity \citep{2011ApJ...735L..46B}. In addition, the
abundance trends found in the open cluster population appear to follow
the abundances found in the thin disk very well, as shown in
\citet{2010AJ....139.1942F} where they compare the [O/Fe] as a
function of [Fe/H] with the data for kinematically selected thin and
thick disk stars \citep{2004A&A...415..155B}.

Over the last few decades several studies and small and large surveys
of both globular and open clusters in the Milky Way have been
undertaken.  In spite of all these efforts, there is still scope for
very large and comprehensive surveys of stellar clusters. For example,
only about 10\% of the known $\sim$1800 open clusters have metallicity
estimates available in the literature \citep{2010IAUS..266..106M}.
While about half of them have distances, ages, and reddening
estimated. For the globular clusters the situation is somewhat more
robust with 1/3 of the $>$150 known globulars having metallicity
estimates. Large area coverage not only in photometry but also in spectroscopy
is clearly desirable.

\section{Clusters and the underlying field populations}
\label{sect:field}

While the globular clusters appear to essentially trace the spheroidal
components of the Milky Way, the Galactic bulge and halo, the open
clusters reside in the disk. Essentially all open clusters have
$|b|<20^{\circ}$ \citep[see, e.g., Figure 1
in][]{2010IAUS..266..106M}.  Thus the globulars trace the
older parts, both the metal-poor halo and the metal-rich bulge, of
the Milky Way, whilst the open clusters trace the, younger, stellar disk. The
question then arise -- Are these just ``spatial coincidences'' or do
the cluster systems truly trace the underlying field population and
can they tell us something about how those stellar components of the
Milky Way formed and evolved?

It might first be interesting to consider if the populations of
stellar clusters as we know them today are representative samplings of
the underlying, full cluster populations. The answer to this is
related to how we discover stellar clusters. For example, it is much
easier to find a stellar system in parts of the sky that are less
crowded and/or have low extinction. The stellar disk has, especially
at the very lowest latitudes, a very large extinction, reaching many
magnitudes towards the central parts of the Galaxy. How the presence
of extinction influences our ability to find and study open clusters
in the disk is illustrated in Fig.\,2 in \citet{2010IAUS..266..106M}
which shows how the known open clusters nicely traces the low
extinction areas in the first quadrant. Photometric searches in near
infra-red surveys, such as 2MASS, have revealed many clusters hidden
to surveys in the optical \citep[e.g.,][]{2007MNRAS.374..399F}. The
need for spectroscopic follow-up to confirm clusters against asterisms
is further discussed in \citet{2011A&A...530A..32B}.  The photometric
near infra-red VVV survey (see page\,\pageref{vvv}) has already
discovered 96 new infra-red open clusters and stellar groups.  These
cluster candidates are mainly faint and compact and are younger than
5\,Myr \citep{2011arXiv1106.3045B}. The high extinction of these new
clusters, up to 20 mag in $V$, show how important it is
to have near infra-red surveys to fully explore the cluster population
that resides in the plane.  In addition, for open clusters there is a
clear tendency that more detailed studies have concentrated on the
brighter or more well-populated clusters where the colour-magnitude
diagrams are better populated \citep{2010IAUS..266..106M}. In summary,
we have only begun to scrape the surface of the properties of the open
cluster system. Nevertheless, the open clusters, which clearly are
currently forming within the stellar disk, are more likely directly
related to the field population as such than the globulars are. It is
worth pointing out that also globular clusters are found in the near
infra-red surveys \citep[e.g.,][]{2008A&A...479..741B}. It will be
interesting to see if the gap in, e.g., age and mass between the two
sets of clusters will diminish or remain thanks to the new clusters
found.

As discussed earlier, some globular clusters are likely accreted
from other galaxies and are thus fossil records of other stellar
populations than those in the Milky Way. It is in this
context that good elemental abundances become interesting. We expect
the abundance trends or patterns to be different for different systems
\citep{2009ARA&A..47..371T}.  This has, for example, been used in a
wide range of studies of the Sagittarius dSph, its associated stellar
clusters, and streams to pin down their common origin
\citep[e.g.,][]{2004AJ....127.1545C,2005A&A...437..905S,2007A&A...465..815S,2007A&A...464..201M,2008AJ....136..731M}. 

Stars form in cluster of varying sizes. Lighter clusters are
anticipated to dissolve into the surrounding field on a short
timescale. Olin Eggen was perhaps the first to investigate the so
called moving groups \citep[see,
e.g.,][]{1996AJ....112.1595E,1966IAUTB..12..432E}. He identified a
number of such groups based on their common space motions. Improvement
in data, in particular the advent of the Hipparcos catalogue
\citep{1997ESASP1200.....P} which enabled the calculation of reliable
space velocities, has re-ignited the interest in this field. A number
of moving groups have been re-assessed and several new ones have been
found \citep[examples
include][]{2010A&ARv..18..567K,2010AstL...36...27B,2007A&A...461..957F,2006A&A...449..533A,1998A&A...339..831B}. However,
dynamical investigations indicate that some of the moving groups are
kinematic features rather than dissolved stellar clusters
\citep{2000AJ....119..800D}.  If we assume that the stars were
actually born together then they should not only have a common age but
also share the abundance pattern. Thus spectroscopic abundance
analysis to obtain elemental abundances is necessary to check if it is
a cluster or a dynamical feature. \citet{2007AJ....133..694D} showed
that HR\,1614 is indeed a dissolved cluster with the stars sharing a
common abundance pattern. The Hercules stream on the other hand was
shown to just sample both thin as well as the thick disk abundances
and ages, with a rather broad distribution in [Fe/H] proving its
origin to be, most likely, the local dynamical effects of the bar
\citep{2007ApJ...655L..89B}. The prospects of {\sl Gaia} to improve
our understanding of the stirring by the bar and spiral arms, and the connection to 
moving groups and dissolving stellar clusters
in terms of Galactic evolution is further discussed in, e.g.,
\citet{2011arXiv1106.1170A} and \citet{2010ApJ...725.1676B}.

Also globular clusters loose stars which populate the field. Prominent
streams have been detected, e.g., for Palomar\,14 and 5
\citep[][respectively]{2011ApJ...726...47S,2001ApJ...548L.165O}.

\section{Surveys to complement and enhance Gaia}
\label{sect:gaia}

{\sl Gaia}\footnote{{\sl Gaia}'s science performance is available at \\
  \href{http://www.rssd.esa.int/index.php?project=GAIA&page=Science_Performance}{\tt
    http://www.rssd.esa.int/index.php?project=GAIA\&page=Science\_Performance}. The
  numbers on these web-pages are predicted to be robust until the
  mission flies, and no further updates will be given.}  will observe
a billion objects down to $G\simeq 20$\footnote{{\sl Gaia} magnitudes
  are in the white-light $G$-band which covers
  $\sim330-1050$\,nm.}. {\sl Gaia} will, apart from parallaxes and
proper motions, provide photometric information for all objects
enabling astrophysical classification (e.g., star or quasar) and
astrophysical classification (e.g., effective temperatures,
photometric redshift). For the 150 million stars brighter than
$G\simeq 16$ the on-board spectroscopic instrument will in addition
provide radial velocities. For even brighter stars, $G\simeq 12$ which
corresponds to about 5 million stars, the spectrograph will in
addition give interstellar reddening, atmospheric parameters, and
rotational velocities. It will provide elemental abundances for tho
two billion brightest stars ($G\simeq 11$). This leaves almost 90\% of
the stars observed by {\sl Gaia} without radial velocity estimates
and even more stars without any additional astrophysical
information. Such information can be got from ground-based
observations using efficient, multi-fibre spectrographs on 4- and
8-meter class telescopes.

That {\sl Gaia} needs to be complemented with ground-based
observations has long been recognized. For example the ESA-ESO Working
group on galactic populations, chemistry and dynamics set out the
challenges and priorities already in their report 2008
\citep{2008ewg4.rept.....T,2008Msngr.134...46T}, where they especially recognized the
importance of powerfully multi-object spectrographs with high
multiplex. ASTRONET's\footnote{ASTRONET was created by a group of
  European funding agencies in order to establish a long-term roadmap
  for European astronomy. The report {\sl Science Vision for European
    astronomy} is available at
  \href{http://www.astronet-eu.org/-Science-Vision-}{\tt
    http://www.astronet-eu.org/-Science-Vision-}.} consultations 
reach them same conclusions and their report concluded: ``It is crucial
to supplement the Gaia data-set with dedicated ground-based
spectroscopic programmes, in order to obtain the radial velocity and
detailed chemical abundances for fainter stars.''

Depending on what scientific questions that are being asked the
requirements on the derived abundances will differ. If for example we
want to figure out which clusters belong to the Sagittarius dSph
galaxy we may only need an internal precision of about 0.1-0.2\,dex,
while if we want to identify individual stars that belong to a
dispersed stellar cluster we may need an internal precision in our
measurements of about 0.05\,dex or less. The latter can readily be
refereed to as the chemical tagging suggested by
\citet{2002ARA&A..40..487F} while the former is better refereed to as a
chemical labelling, a term introduced by Vanessa Hill. It is important
to make these distinctions because what you can learn from the various
types of measurements differs significantly and has a straightforward,
direct impact on the design of stellar spectroscopic surveys. Very
high precision within a study is often reached by studying stars that
in all respects are very similar in their stellar parameters and only
differ in age and elemental abundances. Only selecting similar types
of stars for the survey is a technique that has been successfully
employed many times. Good examples include
\citet{1993A&A...275..101E}, \citet{2011MNRAS.414.2893F}, 
\citet{2004A&A...415..155B}. Recently, the same methodology has been used 
to identify solar twins \citep{2010A&A...522A..98M}.

A summary of the necessary requirements on the equipment, based on
actual science cases, was set out by a working group inside
GREAT-ESF\footnote{GREAT-ESF is the Gaia Research for European
  Astronomy Training. Funding from \href{http://www.esf.org/}{ESF}
  provides for meetings and visits. Its web-site is
  \href{http://www.ast.cam.ac.uk/ioa/GREAT/}{\tt
    http://www.ast.cam.ac.uk/ioa/GREAT/} and its wiki is at
  \href{http://camd08.ast.cam.ac.uk/Greatwiki/GreatHome}{\tt
    http://camd08.ast.cam.ac.uk/Greatwiki/GreatHome}}. In the
resulting document\footnote{The latest version,April 2010, is
  available
  \href{http://camd08.ast.cam.ac.uk/Greatwiki/GreatCds?action=AttachFile&do=get&target=gcds-case-v0.9.pdf}{here}}
we concluded that the best synergies with the on-board instrumentation
on {\sl Gaia} would be provided with the following three broad sets of
instruments:

\begin{enumerate}
\item Low resolution spectroscopy for completion of the $6D$ phase space
  information for stars with $16 < V < 20$.
\item $R = 20000$ spectroscopy of metal-poor disk and halo stars, giants
  as well as dwarfs, and stars in nearby dwarf galaxies. This mode
  will enable observations of stars at very large distances (4-m class
  telescopes).
\item $R = 40 000 - 60 000$ multi-fibre and perhaps single slit
  spectroscopy of selected populations of metal-rich stars, e.g. disk
  and (outer) bulge (4- and 8-m class telescopes).
\end{enumerate}

It is now very exciting times when some of these thoughts are 
put into practise through a number of  efforts. Below some 
of the most relevant projects are being discussed.

\subsection{The Gaia-ESO Survey}

The {\sl Gaia-ESO Survey} is the result of the community's response to
ESO’s call for public spectroscopic surveys. The final proposal was
for 3000 hours on FLAMES-UVES on VLT/UT2 to obtain radial velocities
and elemental abundances for $>10^5$ stars and $>100$ stellar clusters
covering all the major stellar components of the Milky Way.  It has a
fully stand-alone science case and will provide the first homogeneous
overview of the distributions of kinematics and elemental abundances
in the Milky Way. It is also designed to, later, take advantage of the
{\sl Gaia} astrometry. 

The FLAMES-UVES spectrographs provides a unique opportunity to probe
the Galactic components both locally in very high detail as well as on
a larger scale with still very good elemental abundances.  The FLAMES
spectra will enable the determination of individual elemental
abundances in each star, yield precise radial velocities for a $4D$
kinematic phase-space, and map both kinematic gradients as well as
abundance -- phase-space structure throughout the Galaxy.  The
high-resolution fibres in the UVES spectrograph will be used to obtain
high signal-to-noise spectra for a few thousand dwarf stars within
2\,kpc providing a complete census of the distribution functions for
the elemental abundances present in the old disk.

The {\sl Gaia-ESO Survey} will be the first homogeneous spectroscopic
survey of a statistically significant sample of stellar clusters.
Encompassing clusters with ages from $10^6$ up to $10^9$ years, in
different environments, with different richnesses and cluster masses,
and different galacto-centric positions.  For each cluster the survey
will provide a ``complete'' stellar sample based on
detailed chemistry as well precise kinematics. In addition it will
also provide measures of stellar activity, quantitative mass-loss
estimates for early-type stars, and refined memberships for cluster
members.

An important aspect of this survey is the homogeneous abundance scale for
stellar cluster and field stars. Thus enabling a truly differential study
between the field the clusters are embedded in and the clusters themselves. 
Combined with the accurate velocity determinations for the stellar clusters
this will lead to a deeper understanding of how the field is populated with
stars from dispersing stellar clusters.

The {\sl Gaia-ESO Survey} includes more than 300 astronomers across Europe
and is PIed by Gerry Gilmore and Sofia Randich. The survey will have its own
dedicated web-space at {\tt www.gaia-eso.eu}.

\subsection{Selected  instruments and their associated surveys}

Currently, a number of multi-fibre spectrographs are either being
commissioned, being built, or are undergoing studies. All of these are
of direct relevance to the {\sl Gaia} mission.  In addition, many
photometric surveys provide valuable data that will enhance the {\sl
  Gaia} results. For instance, SkyMapper \footnote{More about the
  project, including the filter curves, can be found at
  \href{http://msowww.anu.edu.au/skymapper/}{http://msowww.anu.edu.au/skymapper/}}
has a filter system that is designed to be sensitive to stellar
metallicity and gravity. This will complement the {\sl Gaia} data with
photometric metallicities for all stars too faint to get good
metallicities from the on-board instruments. The SkyMapper filters
roughly follow the Str\"omgren photometric system for which good
metallicities can be derived for a range of stellar evolutionary
stages \citep[see, e.g., Ad\'en et
al. submitted,][]{2011A&A...530A.138C,2011A&A...525A.153A,2010A&A...521A..40A}. We
may also foresee important complementarity to the {\sl Gaia} from the
photometric VISTA surveys, already operational and including the VISTA
Variables in The Via Lactea (VVV\label{vvv}) \footnote{The survey wiki
  is available at
  \href{http://mwm.astro.puc.cl/mw/index.php/Main_Page} {\tt
    http://mwm.astro.puc.cl/mw/index.php/Main\_Page}} multi-epoch
public survey which will build a high resolution $3D$ map of the
Galactic bulge including stellar variability
\citep{2010NewA...15..433M}. The Galactic bulge is notoriously
difficult to study thanks to the high and variable reddening. Thus the
VVV will be a necessary complement to {\sl Gaia}, which is not
optimized for studies in such crowded regions with very high
extinction. The variability studies will also add an interesting time
dimension, enabling asteroseismology to be carried out on red
giants. With such data also red giant branch stars can have their ages
determined, something that is impossible without the very accurate
masses provided by asteroseismology.  Of the current surveys
the Sloan Digital Sky Survey (SDSS) is probably that with the largest
impact so far. The combination of its large photometric data-base and
basic stellar parameters derived from low-resolution spectra
\citep[see, e.g.,][]{2011arXiv1104.3114L,2008AJ....136.2022L} with the
{\sl Gaia} parallaxes and proper motions will very quickly result in
new insights into the formation and evolution of the Milky Way. Other
photometric surveys of interest includes Pan-STARSS\footnote{More
  information as its surveys progress can be found at
  \href{http://pan-starrs.ifa.hawaii.edu/public/}{\tt
    http://pan-starrs.ifa.hawaii.edu/public/}} and Large Synoptic
Survey Telescope \citep[LSST][]{2008arXiv0805.2366I}. LSST can be
thought of as a deep extension {\sl Gaia}.  {\sl Gaia}'s error in
proper motion at $r\sim 19$ is 0.1 mas\,yr$^{-1}$. This is LSST's
smallest error and it performs to the same precision down to about
$r=21$ \citep[see further discussions on the synergy between {\sl
  Gaia} and LSST in, e.g.,][]{2011EAS....45..281J}.

\paragraph{RAVE} (the Radial Velocity Experiment) is an on-going
spectroscopic survey obtaining radial velocities with an accuracy of about 2
km\,s$^{-1}$ for up to a million stars. The spectra cover, as for {\sl
  Gaia}, the Ca\,{\sc ii} triplet region.  Elemental abundances for
the RAVE stars have been derived from these as well as from follow-up
spectra
\citep[e.g.,][]{2010ApJ...721L..92R,2010ApJ...724L.104F}. Other
recent, interesting results, showing the value of large, comprehensive
spectroscopic data-sets, concern, e.g., the detection of young, moving
groups \citep{2011MNRAS.411..117K}. Further examples can be found in
the publication lists and news items on their
web-page\footnote{\href{http://www.rave-survey.aip.de/rave/}{\tt
    http://www.rave-survey.aip.de/rave/}}.

\paragraph{APOGEE}\label{par:apogee} is part of SDSS-III\footnote{\href{http://www.sdss3.org/}{\tt http://www.sdss3.org/} and \href{http://www.sdss3.org/surveys/apogee.php}{\tt http://www.sdss3.org/surveys/apogee.php}}, which also includes the 
Baryon Oscillation Spectroscopic Survey (BOSS) as well as the MARVELS
search for exo-planets, and the Sloan Extension for Galactic
Understanding and Exploration 2 (SEGUE-2). The survey starts 2011. The
spectrograph has 300 fibres, a wavelength coverage of $1.52
-1.69$$\mu$m, and a resolution of about 20\,000. Over a period of four
years it will obtain spectra (S/N=100) for 100\,000 red giant stars
down to $H=13.5$ selected from 2MASS. From these spectra abundances
for more than 15 elements will be derived as well as velocities with
errors on the order of 0.5\,km\,s$^{-1}$. The survey will observe
around 200 stellar clusters, including 17 calibrating cluster. Of the
calibrating clusters 12 are open clusters and 5 globular clusters
\citep{2010jena.confE.136F}. APOGEE is planning to work together with
CoRoT and Kepler (KASC) to combine the good mass estimates for red
giant branch stars provided by the space missions from
asteroseismology with the detailed elemental abundances from
APOGEE. The accuracy of the mass estimate is improved by the good
abundance determinations. Combined, they will yield reliable ages for
stars on the red giant branch. A task that otherwise is virtually
impossible due to the very closely packed isochrones in this
evolutionary phase.

\paragraph{HERMES} \label{hermes} is a multi-fibre spectrograph to
operate on the the 3.9 meter Anglo-Australian telescope (AAT). The
instrument will use the existing 2dF optical fibre positioner to place
the 400 fibres over the two-degree field of view. The Galactic
Archeology with HERMES (GALAH
\footnote{\href{http://www.aao.gov.au/HERMES/GALAH/Home.html}{\tt
    http://www.aao.gov.au/HERMES/GALAH/Home.html}. A useful
  presentation is also available at
  \href{http://www.aao.gov.au/HERMES/ScienceWorkshop/Talks/gds.pdf}{\tt
    http://www.aao.gov.au/HERMES/ScienceWorkshop/Talks/gds.pdf}.})
project starts in 2013. Observations are carried out in four
wavelength ranges in the visible covering 25 different species,
including all the major nucleosynthetic channels. The resolution is
relatively high at 30\,000 with an option for even higher
resolutions. The survey is limited to stars brighter than $V$=14. Even
so, it will provide a very important first ``all-sky'' complement to
{\sl Gaia} delivering high quality elemental abundances for almost a
million stars without elemental abundances directly measured using the
{\sl Gaia}-spectra. GALAH will use open and globular clusters for
calibrating purposes as well as some very metal-poor stars to provide
calibrations below the canonical --2.2\,dex, where the globulars stop.

\paragraph{4MOST} is a proposed multi-object fibre spectrograph with a
very high multi-plex to go on one of ESO's 4-meter class
telescope. The object is to provide {\sl Gaia} and {\sl eROSITA} with
the necessary ground-based spectroscopy. The baseline is for a 1500
fibres with 3 degree$^{2}$ field-of-view with a goal of 3000 fibres
over a 5 degree$^{2}$ field-of-view. The spectrograph has a low
resolution mode for obtaining radial velocities and rough stellar
parameters and a high resolution mode for chemical labelling. Examples
of the Milky Way science that will be done with 4MOST include detailed
studies of moving groups, dynamical structures as well as dissolving
clusters, out to about 10\,kpc. Hipparcos data allows us to study
these features out to about 200\,pc and does not allow for strong
constraints on the bar and spiral arms
\citep{2011arXiv1106.1170A,2010MNRAS.407.2122M}. Its high-resolution
survey will obtain chemo-dynamical data for more than 10$^{6}$ stars
which allows, e.g., for studies of radial abundance gradient, i.e. the
build up of the stellar disk, radial migration, formation of the thick
disk etc. All open clusters will be thoroughly covered and the
examples given above will allow for an even deeper understanding of
the connection between the field and the clusters. A sub-survey aims
at looking for truly large samples of the most metal-poor stars
known. The instrument is undergoing a detailed phase A study and is
PI:ed by Roelof de Jong (AIP)\footnote{A presentation of the many
  science goals and the various design concepts was given at the
  GREAT-ESF meeting in Brussels 2011. It is available
  \href{http://camd08.ast.cam.ac.uk/Greatwiki/GreatMeet-20110621?action=AttachFile&do=get&target=4MOST_Brussels_GREAT11.pdf}{here}.}.

\paragraph{MOONS} is a proposed multi-fibre spectrograph operating in
the near infra-red on the VLT employing 500 fibres over a
field-of-view of 500 arcmin$^{2}$. The science goals include {\sl
  Gaia} follow-up and several extra-galactic science cases. From the
point of view of Galactic Archeology and {\sl Gaia} its major
advantage over many other instruments is the combination of
observations in the near infra-red with an 8-meter telescope. Thus
it will provide a necessary complement to, e.g., {\sl Gaia}'s radial
velocities in the inner disk and Galactic bulge region.  MOONS is also
the ideal instrument to do spectroscopic follow-up and confirmation of
cluster candidates found by photometric surveys in the near infra-red,
as exemplified by the new results from the VVV survey
\citep{2011arXiv1106.3045B}. The instrument is undergoing a detailed
phase A study and is PI:ed by Michele Cirasuolo (Edinburgh).

\paragraph{WEAVE} is a proposed multi-fibre spectrograph for the 4
meter William Herschel Telescope on La Palma. The design includes
about 1000 fibres within a 2$^{\circ}$ field of view at the lower
resolution of 5000 to, from the point of view of {\sl Gaia} follow-up,
provide radial velocities with an accuracy of less than 5 km\,s$^{-1}$
for stars with $17<V<20$ and a high-resolution mode with R=20\,000 for
abundance determinations for stars
\citep{2010SPIE.7735E.242B}\footnote{The project's web-page is at
  \href{http://www.ing.iac.es/weave/}{\tt
    http://www.ing.iac.es/weave/}}. The instrument would, e.g., be
ideal to characterize halo streams through chemical labelling, where a
relatively low surface density maps well to the field of view and
fibre density.

\paragraph{Guoshoujing Telescope}(formerly LAMOST\footnote{Their
  official web-site is at \href{http://www.lamost.org/website/en}{\tt
    http://www.lamost.org/website/en}}) has a 5$^{\circ}$ field of
view and 4000 optical fibres. It is currently working in a low
resolution mode but updates to a higher resolution is foreseen. It
covers the wavelength regions of 370-590\,nm and 570-900\,nm. The
Guoshoujing Telescope telescope has its own, proposed, open cluster
survey, LOCS \citep{2009IAUS..254P..15C}, to, e.g., look at abundance
gradients in the Galactic disk.

\section{Concluding remarks}
\label{sect:final}

Although it is reasonably feasible to compare the major properties of
a Milky Way galaxy formed in a $\Lambda$CDM simulation with the
overall results from current photometric and spectroscopic surveys
{e.g., SDSS, SEGUE, RAVE) it is far from straightforward to constrain the models
with the available data,  mainly for
  two reasons: 1) the resolution in the models is too low to allow the
  investigation of the small scale features that we actually see in a
  real galaxies, 2) the selection of targets in the observational 
studies is either not well understood/documented or not optimized for
the particular question of interest. Future surveys, especially the
massive spectroscopic surveys operating after {\sl Gaia} will be able to 
have very good selection criteria combined with, almost, exhaustive coverage
of a given stellar population, hence enabling a more detailed comparison
with models. 

The several examples of past studies of the Milky Way and its stellar
components given in this review show the need for large spectroscopic
surveys to enable a full disentangling of the formation processes
involved in shaping our Galaxy. Indeed, with the combination of {\sl
  Gaia}'s distances and proper motion and a ground based massive
follow-up providing the $6D$ phase-space as well as the
multi-dimensional abundance and age space we will finally be able to
start testing the models for real \citep[compare, e.g., the proposed
tests in][which are only feasible with this new
data]{2009MNRAS.400L..61S}. 
In addition, prepare to find the unexpected. This means,
be ambitious also when it comes to the elemental abundances and aim for abundance
errors smaller than 0.05\,dex (internally) in order to find and tag also the small
building blocks  of our Galaxy, i.e. the open clusters.

%
\small  
%
\section*{Acknowledgments}   
%
I would like to thank the organisers of the excellent workshop
``Stellar Clusters \& Associations: A RIA Workshop on Gaia" for a very
nice meeting and their generous hospitality during the conference.

%
%
%
%
%

\bibliographystyle{aa}
\bibliography{mnemonic,Feltzing}

\end{document}